\documentstyle[prl,aps,twocolumn,amssymb]{revtex}
\begin{document}
\twocolumn[\hsize\textwidth\columnwidth\hsize\csname @twocolumnfalse\endcsname
\draft
\title{Mechanical Mixing in Nonlinear Nanomechanical Resonators}
\author{A. Erbe$^*$, G. Corso$^{**}$, H. Kr\"ommer$^*$,
A. Kraus$^*$, K. Richter$^{**}$, and R.H. Blick$^*$}

\address{$^{*}$ Center for NanoScience and Sektion Physik,
Ludwig-Maximilians-Universit\"at, Geschwister-Scholl-Platz 1,
80539 M\"unchen, Germany. \\}

\address{$^{**}$ Max-Planck-Institut f\"ur Physik komplexer Systeme,
  N\"othnitzer Str. 38, 01187 Dresden, Germany.
        }
\date{\today}
\maketitle

\widetext

\begin{abstract}
Nanomechanical resonators, machined out of Silicon-on-Insulator wafers,
are operated in the nonlinear regime to investigate higher-order
mechanical mixing at radio frequencies, relevant to signal processing
and nonlinear dynamics on nanometer scales.
Driven by two neighboring frequencies the
resonators generate rich power spectra exhibiting a multitude of satellite
peaks.
This nonlinear response is studied and compared to
$n^{th}$-order perturbation theory and nonperturbative numerical calculations.
\end{abstract}
]

\narrowtext

Mechanical devices in combination with modern
semiconductor electronics offer great advantages as for example their
robustness against
electrical shocks and ionization due to radiation. In outstanding work by
Rugar and Gr\"utter~\cite{rugar91} the importance for applications in
scanning probe microscopy of mechanical cantilevers was demonstrated.
Greywall~{\it et al.}~\cite{greywall94} investigated noise evasion techniques
for frequency sources and clocks with microscopical mechanical resonators.
The main disadvantage of mechanical devices so far is the low speed of
operation. This has been overcome with the realization of nanomechanical
resonators,
which allow operation at frequencies up to
500~MHz~\cite{cleland96,carr97,erbe98,pescini99,kroemmer99}.

In the present work we realize such a nanomechanical resonator to study
its nonlinear dynamics and its mechanical mixing properties.
Mixing is of great importance for signal processing in common electronic
circuits. Combining signal mixing with the advantages of mechanical systems,
i.e.\ their insensitivity to the extremes of temperature and radiation,
is very promising, especially when considering the high speed of operation
currently becoming available.
Here we present measurements on such a nonlinear nanomechanical
resonator, forced into resonance by application of
two different but neighboring driving frequencies. We also present a
theoretical
model, based on the Duffing equation, which accurately describes the
behavior of the  mechanical resonator. The model gives insight into the
degree of nonlinearity of the resonator and hence into the
generation of higher-harmonic mechanical mixing.

The starting materials are commercially available 
\hspace{0.1cm} Silicon-on-insulator 
(SOI) substrates with thicknesses of the Si-layer and the 
SiO$_2$ sacrificial layer of 205~nm and 400~nm, 
respectively (Smart-Cut wafers).  \hspace{0.1cm}  The gate leads
connecting the resonator
to the chip carrier are defined using optical lithography. In a
next step the nanomechanical resonator is defined by
electron beam lithography. \hspace{0.1cm} 
The sample is dry-etched in a reactive-ion etcher (RIE) in order to
obtain a mesa structure with clear-cut walls. 
Finally, we perform a hydro-fluoric (HF) wet-etch step
in order to remove the sacrificial layer 
below the resonators and the metallic etch mask. 
The last step of processing is critical point drying, 
in order to avoid surface tension by the solvents. 
\hspace{0.1cm} The suspended resonator is shown in a scanning 
electron beam micrograph in Fig~1(a): The beam has a 
length of $l = 3~\mu$m, a width of $w =
200$~nm, and a height of $h = 250$~nm and is clamped on both sides.  
The inset shows a close-up of the suspended beam. 
 The restoring force of this Au/Si-hybrid beam is dominated by 
the stiffer Si supporting membrane. \hspace{0.1cm} 
The selection of the appropriate HF etch
allows for attacking only the Si and thus the minute determination of
the beam's flexibility and in turn the strength of the nonlinear
response.

The chip is mounted in a sample holder and a small amount of
$^4$He exchange-gas is added (10 mbar) to ensure thermal coupling.
The sample is placed at 4.2~K in a magnetic field, directed in parallel
to the sample surface but perpendicular to the beam. When an
alternating current is applied to the beam a Lorentz force arises
perpendicular to the sample surface and sets the beam into mechanical
motion. For characterization we employ a spectrum analyzer (Hewlett Packard
8594A):
The output frequency is scanning the frequency range of 
interest~($\sim 37$~MHz),
the reflected signal is tracked and then amplified (setup $\alpha$ in
Fig.~1(b), reflectance measured in mV).
The reflected power changes when the resonance condition is met, which
can be tuned by the gate voltages $V_g$ in a range of several 10 kHz.
The mixing properties of the suspended nanoresonators are probed with
a different setup comprising two synthesizers (Marconi~2032 and Wavetek 3010)
emitting excitations at constant, but different,
frequency (setup $\beta$ in Fig.~1(b)). Here, the reflectance is measured in
dBm for better comparison of the driving amplitudes and the mixing products.
The reflected power is finally amplified and
detected by the spectrum analyzer.

In Fig.~2 the radio-frequency (rf) response of the beam near resonance is
depicted
for increasing magnetic field strength $B\!=\!0,1,2,\ldots,12$~T.
The excitation power of the spectrum analyzer was fixed at $- 50$~dBm.
The mechanical quality factor, $Q = f/\delta f$, of the
particular resonator under test in the linear regime is $Q = 2330$.
As seen the profile of the resonance curve changes from a symmetric shape
at moderate fields to an asymmetric, sawtooth shape at large field values,
characteristic of an oscillator operated in the nonlinear regime.

This behavior can be described by the Duffing equation
\begin{equation}
    \ddot x(t) + \mu \dot x(t) + \omega_0^2 x(t) + \alpha x^3(t) = F(t)
\label{duffing}
\end{equation}
with a positive prefactor $\alpha$ of the cubic term being the parameter of
the strength of the nonlinearity\cite{parlitz}.
In Eq.~(\ref{duffing}) $\mu$ is the damping coefficient of the mechanical
system,
$\omega_0 = 2 \pi f_0$, where $f_0$ is the mechanical eigenfrequency of the
beam, and $x(t)$ its elongation. In our case the external driving $F(t)$
is given by the Lorentz force:
\begin{equation}
   F(t) = \frac{l B}{m_{\rm eff}} I(t) = \frac{l B}{m_{\rm eff}} I_{0} \cos
   (2\pi f t) \, ,
\label{lorentz}
\end{equation}
where $l\!=\!1.9\cdot 10^{-6}$~m is the effective length
and $m_{\rm eff}\!=\!4.3\cdot 10^{-16}$~kg is the
effective mass of the resonator. $B$ is the magnetic field and
$I_0$ the input current corresponding to the amplitude of the driving power.

Solving Eq.~(\ref{duffing}) and computing the amplitude of the oscillation
as a
function of the driving frequency $f$ for several excitation strengths
reproduces the measured curves shown in Fig.~2.
The solutions at large power exhibit a region where three
different amplitude values coexist at a single frequency.
This behavior leads to a hysteretic response in the measurements
at high powers (e.g.\  $-50$~dBm)\cite{kroemmer99},
as shown in the inset of Fig.~2, where we used an
external source~(Marconi) to sweep the frequencies in both directions.
If the frequency is increased (inverted triangles ($\bigtriangledown$) in the
inset), the resonance first follows the
lower branch, and then suddenly jumps to the upper branch.
When sweeping downwards from
higher to lower frequencies (triangles ($\bigtriangleup$)),
the jump in resonance occurs at a different frequency.

Turning now to the unique properties of the nonlinear nanomechanical
system: By applying two separate frequency sources as sketched in
Fig.~1(b) (setup $\beta$) it is possible to demonstrate mechanical
mixing, as shown in Fig.~3(a). The two sources are tuned to
$f_1 = 37.28$~MHz and $f_2 = 37.29$~MHz with  constant offset
and equal output power of $-48$~dBm, well in the nonlinear regime. Without
applying a magnetic field the two input signals are simply reflected
(upper left panel). Crossing a critical field
of $B\simeq 8$~T higher-order harmonics appear. Increasing the field
strength further a multitude of satellite peaks evolves. As seen the limited
bandwidth of this mechanical mixer allows effective signal filtering.

Variation of the offset frequencies leads to the data presented
in Fig.~3(b): Excitation at $- 48$~dBm and $B = 12$~T with the
base frequency fixed at $f_1 = 37.290$~MHz and varying the sampling frequency
in 1~kHz steps from $f_2=37.285$~MHz to 37.290~MHz yields satellites
at the offset frequencies
$f_{1,2} \pm n\Delta f$, $\Delta f =  f_1 - f_2 $. The dotted line is taken at
zero field for comparison, showing only the reflected power when
the beam is not set into mechanical motion. At the smallest offset
frequency of 1~kHz the beam reflects the input signal as a
broad band of excitations.


We model the nanomechanical system as a
Duffing oscillator (\ref{duffing}) with a driving force
\begin{equation}
F(t) = F_1 \cos (2\pi f_1 t) + F_2 \cos (2\pi f_2 t) \, ,
\label{duffingexc}
\end{equation}
with two different, but neighboring, frequencies $f_1$ and $f_2$ and
amplitudes $F_i=l\>B \>I_i/m_{\rm eff}$.

Before presenting our
results of a numerical solution of Eq.~(\ref{duffing})
for the driving forces (\ref{duffingexc})
we perform an analysis based on $n^{th}$-order
perturbation theory\cite{nayfeh} to explain the generation of higher
harmonics. Expanding
\begin{equation}
    x = x_0 + \epsilon x_1 + \epsilon^2 x_2+ \ldots \, ,
\label{expansion}
\end{equation}
where we assume that the (small) parameter $\epsilon$ is of order
of the nonlinearity $\alpha$, and inserting this expansion
into Eq.~(\ref{duffing}) yields
equations for the different orders in $\epsilon$.
In zeroth order we have
\begin{equation}
  \ddot x_0 +\mu \dot x_0 +\omega_0^2 x_0 = F_1 \cos (2\pi f_1 t)
 + F_2 \cos (2\pi f_2 t) \, ,
\label{first}
\end{equation}
to first-order
$ \ddot x_1 +\mu \dot x_1 +\omega_0^2 x_1 + \alpha x_0^3 = 0, $
and similar equations for higher orders. After inserting the solution of
Eq.~(\ref{first}) into the first-order equation and assuming
$f_1 \approx f_2 \approx f_0 = \omega_0/2\pi$,
two types of resonances can be extracted:
One resonance is located at $3 f_0$ which we, however,
could not detect experimentally\cite{bandwidth}.
Resonances of the other type are found at frequencies $f_i \pm \Delta f$.
Proceeding along the same lines in second-order perturbation theory
we obtain resonances at $5 f_0$ and $f_i \pm 2 \Delta f$.
Accordingly, owing to the cubic nonlinear term, $n^{th}$-order resonances
are generated at $(2n+1)f_0$ and $f_i \pm n \Delta f$.
While the $(2n+1) f_0$-resonances
could not be observed, the whole satellite family
$f_i \pm n \Delta f$ is detected in the experimental power spectra
Fig.~3(a,b).

The perturbative approach yields the correct peak positions and, for
$B < 4$~T, also the peak amplitudes.
However, in the hysteretic, strongly nonlinear regime a
nonperturbative numerical calculation proves necessary to explain
quantitatively the measured peak heights.
To this end we determined the parameters entering into Eq.~(\ref{duffing})
in the following way: The damping is estimated from the quality factor
$Q = 2330$ which gives $\mu=50265$~Hz.
The eigenfrequency is $f_0 = 37.26$~MHz as seen from Fig.~2 in the linear
regime.
The nonlinearity $\alpha$ is estimated from the shift\cite{nayfeh}
\begin{equation}
    \delta f(B) = f_{\rm max}(B) -f_0 = \frac{3 \alpha [\Lambda_0(B)]^2}
{32 \pi^2 f_0}
\label{deltaefe}
\end{equation}
in frequency $f_{\rm max}$ at maximum amplitude in Fig.~2. In zero
order the displacement of the beam is given by
$\Lambda_0 = l I_0 B/(4\pi f_0 \mu\  m_{\rm eff})$.
Relation (\ref{deltaefe}) yields with $I_0\!=\!1.9\cdot 10^{-5}$A
a value of $\alpha=9.1\cdot  10^{28}\>({\rm ms})^{-2}$.

We first computed $ x(t)$ by numerical integration of the Duffing equation
with driving (\ref{duffingexc}) and $F_1\!=\!F_2\!=l B I_0/m_{\rm eff}$,
$I_0 = 2.9\cdot 10^{-5}$A. We
then calculated the power spectrum from the Fourier transform $\hat{x}(\omega)$
of $x(t)$ for large times (beyond the transient regime).
For a direct comparison with the measured power $P$ in Fig.~3 we
employ $P \simeq R I_{\rm imp}^2$. Here $R$ is the resistance of the
electromechanical
circuit and $I_{\rm imp} = [4\pi f_0\mu\ m_{\rm eff}/(l B)]\hat{x}(\omega)$
in close analogy to the zero-order relation between displacement $\Lambda_0$
and $I_0$.

The numerically obtained power spectra are displayed in Fig.~4:
(a) shows the emitted power for the same parameters as in Fig.~3(a),
but with $B=4,8,9,10,11$, and $12$~T.
Corresponding curves are shown in Fig.~4(b) for fixed $B$ and
various $\Delta f$ for the same set of experimental
parameters as in Fig.~3(b).
The positions of the measured satellite peaks, $f_i \pm n \Delta f$,
{\em and} their amplitudes are in good agreement with the numerical
simulations
for the entire parameter range shown.
Even small modulations in the peak heights to the left of the two
central peaks in Fig.~3(b) seem to be reproduced by the
calculations in Fig.~4(b).
(Note that the height of the two central peaks in Fig.~3 cannot
be reproduced by the simulations,
since they are dominated by the reflected input signal.)

The numerical results in Fig.~4(a) show clearly the evolution
of an increasing number of peaks with growing magnetic field, i.e.\
increasing driving amplitude. As in the experiment,
the spectra exhibit an asymmetry in
number and height of the satellite peaks which switches from lower
to higher frequencies by increasing the magnetic field from
8~T to 12~T. This behavior can be understood from Eq.~(\ref{deltaefe})
predicting a shift $\delta f$ in resonance frequency with increasing
magnetic field.
This shift is reflected in the crossover in Figs.~3(a) and 4(a).
For $B\! =\! 8$~T the amplitudes of the satellite peaks
are larger on the left than on the right side of the two central peaks.
As the field is increased the frequency shift drives the right-hand-side
satellites into resonance increasing their heights.

The power spectra in Fig.~3(a) and 4(a) are rather insensitive to changes in
magnetic
field for $B < 8$~T compared to the rapid evolution of the
satellite pattern for 8~T~$\!<\! B\! <\! 12$~T. Our analysis shows that this
regime corresponds to scanning through the hysteretic part (inset Fig.~2)
in the amplitude/frequency (or amplitude/$B$-field) diagram, involving
abrupt changes in the amplitudes.
The resonator studied is strongly nonlinear but not governed by chaotic
dynamics.
Similar setups should allow for entering into the truly chaotic regime.

In summary we have shown how to employ the nonlinear response of a strongly
driven nanomechanical resonator as a mechanical mixer in the radio-frequency
regime. This opens up a wide range of applications, especially for signal
processing.
The experimental results are in very good agreement with
numerical calculations based on a generalized Duffing equation, a
prototype of a nonlinear oscillator.
Hence these mechanical resonators allow for studying nonlinear, possibly
chaotic dynamics on the nanometer scale.

We thank J.P.~Kotthaus for helpful discussions.
We acknowledge financial support by the Deutsche Forschungsgemeinschaft (DFG).

\newpage

\begin{figure} [ht]
\begin{center}
\caption[fig1]
{
(a) Scanning electron beam micrograph of the electromechanical
resonator with a length $l = 3~\mu$m, width  $w = 200$~nm,  and height
$h = 250$~nm. The Si-supporting structure is covered
by a thin Au-sheet~(50~nm thick); the electrodes on the left and right
allow tuning of the elastic properties. Inset shows a magnification of the
beam.
(b) Experimental setup for sampling the mechanical properties of the suspended
beam: For characterization we employ a spectrum analyzer scanning the
frequency range of interest ($\alpha$). Mechanical mixing is analyzed
by combining two synthesizers ($f_1$, $f_2$) and detecting the
reflected power ($\beta$).
}
\label{fig1}
\end{center}
\end{figure}

\begin{figure} [ht]
\begin{center}
\caption[fig2]
{
Characterization of the nonlinear response of the suspended beam
by ramping the magnetic field from 0~T up to 12~T,
obtained with the spectrum analyzer operated with output power
level of $-50$~dBm (setup $\alpha$).
Inset shows the measured hysteresis:
$\bigtriangledown$ correspond to an
increase in frequency and $\bigtriangleup$ represent the lowering
branch.
}
\label{fig2}
\end{center}
\end{figure}

\begin{figure} [ht]
\begin{center}
\caption[fig3]
{
(a) Two synthesizers (setup $\beta$ in Fig.~1(b))
running at frequencies of $f_1 = 37.28$~MHz and $f_1 = 37.29$~MHz with
constant
offset (output power $-48$~dBm) induce higher-order harmonics
as a result of mechanical mixing by the nanoresonator
in the nonlinear regime ($B > 8$~T).
(b) Excitation with two frequencies at $- 48$~dBm and $B = 12$~T:
Base frequency is $f_1 = 37.290$~MHz, while the sampling frequency
is varied in 1~kHz steps from $f_2 = 37.285$~MHz to 37.290~MHz. As seen
the spacing of the harmonics follows the offset frequency
$\Delta f =  f_1 - f_2 $.
The dotted line is taken at $B\! =\! 0$~T showing pure reflectance
of the beam without excitation of mechanical motion.
}
\label{fig3}
\end{center}
\end{figure}

\begin{figure} [ht]
\begin{center}
\caption[fig4]
{
Calculation of the power spectra from the numerical
solution of Eqs.~(\ref{duffing}), (\ref{duffingexc}) for the same driving
frequencies as used in Fig.~3.
(a) Variation of magnetic field $B=$4,8,9,10,11, and $12$~T.
(b) Variation of offset frequency at $B=12$~T. Note
that the two central peaks of Fig.~3 are not reproduced by the theory,
since they stem from the reflected input signal.
}
\label{fig4}
\end{center}
\end{figure}


\begin{thebibliography}{99}
\bibitem{bandwidth} The limited sensitivity and bandwidth of the preamplifier
used did not allow us to observe the much weaker
response at the commonly given harmonics $3f_0$ and $f_1 - f_2$.
\end{thebibliography}
\end{document}